\documentclass[12pt,  draftclsnofoot,  onecolumn]{IEEEtran}
\usepackage{amssymb}
\usepackage{graphicx}
\usepackage{amsmath}
\usepackage{epsf}
\usepackage{mathrsfs}
\usepackage{cite}
\usepackage[justification=centering]{caption}

\newtheorem{corollary}{Corollary} 
\newtheorem{lemma}{Lemma} 
\newtheorem{remark}{Remark} 
 
\newtheorem{theorem}{Theorem}
\newtheorem{example}{Example}
\begin{document}
\title{{Perfect Interference Alignment for an Interference Network with General Message Demands}}

\author{\Large Zainalabedin ~Samadi,  
       ~Vahid ~Tabatabavakili and ~Farzan ~Haddadi
\\\small Dept. of Elec. Eng.,   Iran University of Sceince and Technology 
Tehran,   Iran 
\\ \{z.samadi\}@elec.iust.ac.ir
\\ \{vakily,   haddadi\}@iust.ac.ir
}
\maketitle

\begin{abstract}
Dimensionality requirement poses a major challenge for Interference alignment (IA) in practical systems.  This work evaluates the necessary and sufficient conditions on channel structure of a fully connected general interference network to make perfect IA feasible within limited number of channel extensions. So far, IA feasibility literature have mainly focused on network topology, in contrast, this work makes use of the channel structure to achieve total number of degrees of freedom (DoF) of the considered network by extending the channel aided IA scheme to the case of interference channel with general message demands.  We consider a single-hop interference network with $K$ transmitters and $N$ receivers each equipped with a single antenna. Each transmitter emits an independent message and each receiver requests an arbitrary subset of the messages. Obtained channel aiding conditions can be considered as the optimal DoF feasibility conditions on channel structure.  As a byproduct, assuming optimal DoF assignment, it is proved that in a general interference network, there is no user with a unique maximum number of DoF.
\end{abstract}

\section{Introduction}
There are several schemes in multi user networks to manage interference. If interference is weak, the interfering signal is treated as noise. This approach has been used in practice for a long time, e.g., for frequency-reuse in cellular systems. However, information theoretic validation for this
approach has only recently been obtained \cite {Motahari09,   Shang09,   Annapur09}. On the other hand,  for the cases where  interference is strong, the interfering signal can be decoded along with the desired signal and hence canceled \cite{Carleial75,  Sato81,   Han81,   Sankar11,   Sridharan08}. However,  the general condition for strong interference in a $K>2$ user IC is unknown. The problem has been solved for some special cases such as symmetric IC. Lattice-based codes have been used to characterize a “very strong” regime  \cite{ Sridharan08},   the generalized degrees-of-freedom\cite{Jafar10},   and the approximate sum capacity  \cite{ Ordent12},   for symmetric K user ICs.

 If the strength of interference is comparable to the desired signal, then interference is avoided by orthogonalizing the channel access. Primary schemes,   such as time (frequency) division multiple access schemes,   avoid interference by orthogonally assigning the channel between users. Considering the entire bandwidth as a cake,   these schemes cut the cake equally between the  users. Therefore,   if there are $K$ users in the channel,   each user gets roughly $1/K$ of the channel. These orthogonal schemes,   however,   have been proved not to be bandwidth efficient. During the idle condition,   these schemes do not effectively utilize time slot or frequency bandwidth allocated to a user.
 
 In this paper, we explore the regime where all desired and interfering signals are of comparable strength. A recent strategy to deal with interference is interference alignment.  The idea of interference alignment is to coordinate multiple transmitters so that their mutual interference aligns at the receivers, facilitating simple interference cancellation techniques. The remaining dimensions are dedicated for communicating the desired signal, keeping it free from interference. 

 Interference alignment is first introduced by Maddah Ali et. al. \cite{Maddah08},   for X channels. Cadambe and Jafar \cite{Cadam08},   proposed the linear vector interference alignment (LIA) scheme for IC and proved that this method is capable of reaching optimal degrees of freedom of the IC. The optimal degrees of freedom for the $K$ user IC is obtained in the same paper to be $K/2$. The proposed scheme in \cite{Cadam08} is applied over many parallel channels and achieves the optimal degrees of freedom as the signal-to-noise ratio (SNR) goes to infinity. 
 
Nazer et al., \cite{Nazer12},    proposed the so called ergodic IA scheme to achieve $1/2$ interference-free ergodic capacity of interference channel at any signal-to-noise ratio. This scheme is based on a particular pairing of the channel matrices. The scheme needs roughly the same order of channel extension as \cite{Cadam08},    to achieve optimum performance. \cite{Samadi} proposes a new scheme called channel aided IA. It makes use of the channel structure besides the linear IA schemes to achieve total number of DoF in a $K$ user interference channel. In contrast to \cite{Nazer12},   \cite{Samadi} obtains a more general relationship between paired channel matrices,  and thus,  significantly reduces the number of required channel extension. 

A majority of systems considered so far for IA involve only multiple unicast traffic,  where each transmitted message is only demanded by a single receiver. However,  there are wireless multicast applications where  a common message may be demanded by multiple receivers,  e.g.,  in a wireless video broadcasting. The generalization of the multiple unicasts scenario considered in \cite{Cadam08} to the case where each receiver is interested in an arbitrary subset of transmitted messages is considered in \cite{Ke} and DoF region for this network is evaluated in this work. 

In this paper,  we consider the generalization of our previous work \cite{Samadi}, to the case of interference networks with general message demands. In this setup,  there are $K$ transmitters and $N$ receivers,  each equipped with a single antenna. Each transmitter emits a unique message and each
receiver is interested in an arbitrary subset of the messages. Our main result in this paper is the general relationship required  between the paired channel matrices that are suitable for canceling interference,   assuming linear combining of paired channel output signals. 

 So far,  IA feasibility literature have mainly focused on network topology,  using the concept of proper systems \cite{Yetis, Razav, Bresler}. To ease some of interference alignment requirements by using channel structure,   \cite{Leejan09} investigates DoF for the partially connected ICs where some arbitrary interfering links are assumed disconnected. In contrast,  this work evaluates the necessary and sufficient conditions on channel structure of a fully connected general interference network to make perfect  IA feasible within limited number of channel extension. 

The rest of the paper is organized as follows. The system model is introduced In Section 2. In Section 3,   it is argued why linear IA scheme,   over a single antenna interference channel,   can not achieve total number of DoF with limited number of channel extensions. The proposed scheme is described in section 4. Detailed proofs  for our main results are presented in sections 5 and 6.  Concluding remarks are presented in Section 7.

\section{System Model} \label{secsysmod}

\begin{figure}
\centering \includegraphics[scale=0.75]{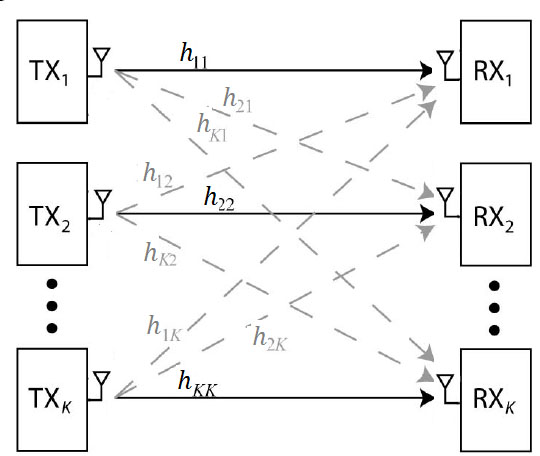}
\caption{K user Interference Channel Model.}
\label{figure:KUser}
\end{figure}

Consider a $K$ user single-hop single antenna interference network. An illustration of system model is shown  in Fig. \ref{figure:KUser}. Each transmitter has one and only one independent message. Each receiver can request an arbitrary set of messages from multiple transmitters.

Let $S_j,  \quad  j=1,  \ldots,  N$ be the set of indices of those transmitted messages requested by receiver $j$ and $ \bar{S}_j$ be the set of indices of those transmitted messages contributing to interference at receiver $j$. Obviously, $S=S_j \cup \bar{S}_j$ is the set of all active transmitters.  All transmitters share a common bandwidth and want to achieve the maximum possible  sum rate along with a reliable communication. Channel output at the $j^{\textrm{th}}$ receiver and over the time slot  $t \in \mathbb{N}$ is characterized by the following input-output relationship:
\begin{eqnarray}
y^{[j]}(t)=h^{[j1]}(t)x^{[1]}(t)+h^{[j2]}(t)x^{[2]}(t) \cdots \nonumber \\+h^{[jK]}(t)x^{[K]}(t)+z^{[j]}(t)
\end{eqnarray}
Where $1 \leq j \leq  N$ is the user index,   $x^{[k]} (t)$  is the transmitted signal symbol of the $k^{\textrm{th}}$ transmitter,    $h^{[jk]} (t), \;   1 \leq k \leq  K$ is the fading factor of the channel from the $k^{\textrm{th}}$ transmitter to the $j^{\textrm{th}}$ receiver  over  $t^{\textrm{th}}$ time slot. We assume that the channel fading factors at different time instants are independently drawn from some continuous distribution.  $z^{[j]}(t)$  is the additive white Gaussian noise at the $j^{\textrm{th}}$ receiver. The noise terms are all assumed to be drawn from a Gaussian independent identically distribution (i.i.d.) with zero mean and unit variance. It is assumed that all transmitters are subjected to  a power constraint $P$:
\begin{eqnarray}
\mathrm{E}(\lVert x_k(t)\rVert ^2) \leq P, \quad  k \in [K],
\end{eqnarray}
where $\mathrm{E}$  is the expectation taken over time, and $[K]$ is defined as $[K]=\{1, \ldots, K\}$. In addition,  the channel gains are bounded between a positive minimum value and a finite maximum value to avoid degenerate channel conditions. Assume that the channel knowledge is causal and available globally,    i.e., over the time slot $t$,    every node knows all channel coefficients $h^{[jk]} (\tau),   \forall j \in [N],   \quad k \in  [K],  \quad  \tau \in \{1,   2,   \ldots,   t\}$. Hereafter,    time index is omitted for convenience.

Ke et al., \cite{Ke}, has  referred to the aforementioned setup as an interference network with general message demands and  has derived the DoF region of this setup. Our objective is to provide necessary and sufficient conditions on channel structure to achieve total number of  DoF using finite channel extension,  assuming  perfect channel state information (CSI) is available at receivers and global CSI at transmitters. Denote the capacity region of such a system as $\mathcal{C}(P)$. The corresponding DoF region is defined as
\begin{eqnarray} \begin{split}
\mathcal{D}=  \{ {\bf d}&=(d_1,  d_2,  \ldots,  d_K) \in \mathbb{R}_+^K: \\ &\exists (R_1(P),  R_2(P),  \ldots,  R_K(P)) \in \mathcal{C}(P), \\ & \quad \quad \textrm{such that}\;  d_k = \lim_{P\rightarrow \infty} \frac{R_k(P)}{\log(P)},  \quad k in [K] \},  
\end{split}\end{eqnarray}

and total number of DoF is defined as $D_s= \max \sum_1^K d_k,  \; \{d_1,  d_2,  \ldots,  d_K\} \in \mathcal{D}$.

\section{Linear IA Limitation} \label{sec3}

Deegrees-of-freedom region for the setup described in section \ref{secsysmod} has been derived in \cite{Ke}  as follows,  
\begin{eqnarray}
\mathcal{D}= \left \{ {\bf d} \in \mathbb{R}_+^K: \sum_{k \in S_j} d_k+\max_{i \in  \bar{S}_j}(d_i) \leq 1,  \; \forall j \in [N] \right \}
\label{dofreg}
\end{eqnarray}
where $\mathcal{S}_j$ is the set message indices requested by receiver $j,  j \in [N]$.

For a single antenna case,  assuming all receivers request the same number of transmitted messages and each transmitter sends message to equal number of receivers,  maximum total number of DoF is $\frac{K}{\beta+1}$,  where $\beta$ is the number of requested messages for each prime receiver,  \cite{Ke}. With prime receiver,  we mean the receivers whose requested message sets are not a subset of any other requested message set. 

Following Theorem describes the only DoF assignment that achieves total number of DoF.
\begin{theorem}
\label{theo1}
The only DoF point that achieves total number of DoF of an interference channel where all receivers request the same number of transmitted messages and each transmitter sends message to equal number of receivers is
\begin{eqnarray}
{\bf d}=\left ( \frac{1}{\beta+1},  \frac{1}{\beta+1},  \ldots,  \frac{1}{\beta+1}\right ).
\label{optdof}
\end{eqnarray}
\end{theorem}
\begin{IEEEproof}
If theorem \ref{theo1} is not true,  there is at least one $d_i,  i=1, \ldots,  K$ which is strictly greater than  $ \frac{1}{\beta+1}$. We would also have the following Lemma. 
\begin{lemma}
\label{lemm1}
In the specified channel structure,  we should have 
\begin{eqnarray}
\max_{i \in  \bar{S}_j}(d_i) \geq \frac{1}{\beta+1},  \quad \forall  j \in [G]
 \end{eqnarray}
 Where $G$ is the number of prime receivers.
\end{lemma}
\begin{IEEEproof}
Assume that there is a $j=j_0$ where $\max_{i \in \bar{S}_{j_0}}(d_i) < \frac{1}{\beta+1}$,  which implies that $d_i < \frac{1}{\beta+1} \quad \forall i \in \bar{S}_{j_0}$.  Thus,  using (\ref{dofreg}),  we will have
\begin{eqnarray} \begin{split}
 \sum_{k \in [K]} d_k&=\sum_{k \in S_{j_0}} d_k+\sum_{k \in \bar{S}_{j_0}}d_k \\ 
 &\leq 1- \max_{i \in \bar{S}_{j_0}}(d_i) +\sum_{k \in \bar{S}_{j_0}}d_k  \\ & < 1+ \frac{K-1-\beta}{1+\beta}=\frac{K}{1+\beta} \\
& \quad \Rightarrow d_{sum}<\frac{K}{1+\beta}.
\end{split}
\label{lemm1eq}
\end{eqnarray}
where $[K]$ is defined as $[K]=\{1, \ldots, K\}$. (\ref{lemm1eq}) is in contrast to the assumption that this DoF assignment achieves total number of DoF,  hence,  the proof of lemma \ref{lemm1}  is complete. 
\end{IEEEproof}

Based on (\ref{dofreg}),  in order to characterize DoF region,  we should consider $G$ inequalities of the form 
\begin{eqnarray}
\sum_{k \in S_j} d_k+\max_{i \in  \bar{S}_j}(d_i) \leq 1,  \quad \forall j \in [G].
\label{dofregineq}
\end{eqnarray}
 Since each message is requested by $G\beta/K$ receivers,  summing all $G$ inequalities,  we have
 \begin{eqnarray}
 G\beta/K \sum_{k \in [K]} d_k + \sum_{j \in [G]} d_{max}^j \leq G
 \label{ineqsum}
 \end{eqnarray}
 where $d_{max}^j$ is defined as $d_{max}^j=\max_{i \in  \bar{S}_j} d_i$. Using the fact that at least there is one $d_{max}^j$ strictly greater than  $ \frac{1}{\beta+1}$,   along with Lemma \ref{lemm1} in (\ref{ineqsum}),  we will have
 \begin{eqnarray} \begin{split}
 & G\beta/K \sum_{k \in [K]} d_k +  \frac{G}{\beta+1}< G \\
& \quad \Rightarrow \sum_{k \in [K]} d_k< \frac{K}{\beta+1}, 
\end{split} \end{eqnarray} 
which contradicts the assumption that this DoF assignment achieves total number of DoF.
\end{IEEEproof}
\begin{figure}
\centering \includegraphics[scale=0.75]{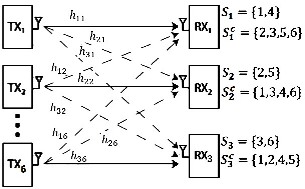}
\caption{$6 \times 3$ user Interference Channel with generalized message set.}
\label{figure:6by3user}
\end{figure}
In the following we will classify interference networks as either regular or irregular based on the optimal number of DoF assigned to each transmitter. Regular networks are the ones whose  only optimal DoF assignment is equal DoF assignment for all active transmitters. Active transmitters are defined as those transmitters with assigned DoF greater than zero. Based on theorem \ref{theo1}, an interference channel where all receivers request the same number of transmitted messages and each transmitter sends message to equal number of receivers, are  regular networks.

\begin{theorem}
\label{theo2}
Assuming channel coefficients to be generic,   total number of the  DoF of an regular network can not be achieved using finite extension of the channel. 
\end{theorem}
\begin{IEEEproof}

Consider an  special case of $6 \times 3$ user interference channel with generalized message set,  the channel structure along with requested set of messages at each receiver is shown in Fig. \ref{figure:6by3user}.  The proof for general case is similar.

We will use the scheme based on \cite{Cadam08} to do interference alignment. Let $\tau$ denote the duration of the time expansion in number of symbols. Here and after, we use the
upper case bold font to denote the time-expanded signals, e.g., ${\bf H}^{[jk]} = \textrm{diag} (h^{[jk]}(1), h^{[jk]}((2), \ldots,h^{[jk]}(\tau))$, which is a size $\tau \times \tau$ diagonal matrix. Denote the beamforming matrix of transmitter $k$ as ${\bf V}^{[k]}$. 

We intend to achieve the outer bound of  $6/3$ DoF for this setup. Considering $3$ extension of this channel. Over this extended channel,  consider a hypothetical achievable scheme where each of the $6$ messages achieves $1$ DoF if possible,  using beamforming at every transmitter and zero-forcing at every receiver. Note that this is the only DoF point in achievable region that achieves total number of DoF of this network, according to theorem \ref{theo1}. 
  
Let message $W^{[j]}$ be beamformed along $3 \times 1$ vector ${\bf V}^{[j]}$ at transmitter $j$. If $j \in S_i$, receiver $i$ intends to decode $W^{[j]}$ using zero-forcing. At receiver $i$,  to decode $2$ independent messages $W^{[j]},  j \in S_i$ using zeroforcing,  the vectors corresponding to the desired messages occupy $2$ linearly independent directions. Since signals come from a space of dimension $3$,  the $4$ interfering vectors must occupy the remaining $1$ dimension. IA requirements can be written as follows.
\begin{itemize}
\item   At receiver $i$,  the vectors ${\bf V}^{[j]},  j \in \bar{S}_i$,   which contribute to interference at receiver $i$,   align within a $1$ demensional subspace,  i.e.,, 
\begin{eqnarray}
\textrm{span}({\bf H}^{[ij]}{\bf V}^{[k]}) = \textrm{span}({\bf H}^{[ik]}{\bf V}^{[k]}), \quad k, j \in \bar{S}_i
\label{SE1}
\end{eqnarray}
Thus,  the total dimension of the interference is $1$ and receiver $i$ can decode all its desired messages.
\end{itemize}

Along with the above conditions,  the desired signal vectors are required to be linearly independent of the interference dimension at each receiver. This requirement implies that,  
\begin{eqnarray}
\mathrm{D}(\textrm{span}[{\bf U}( S_i),  {\bf U( \bar{S}_i)})] = 3, 
\end{eqnarray}
where $\mathrm{D}(S)$ is defined as the dimension of a subspace $S$,  ${\bf U}( S_i)\; \textrm{and} \; {\bf U}( \bar{S}_i)$ are the set of received signal vectors associated with desired and undesired signal vectors,  respectively,  and  $3$ is the total subspace dimension availabe at the receivers. For example,  at user $1$,  ${\bf U}( S_i)$ is obtained as $[{\bf H}^{[12]}{\bf V}^{[1]},  {\bf H}^{[14]}{\bf V}^{[4]}]$. 

This set interference alignment requirements, (\ref{SE1}),  constitue a improper set of equations, \cite{Yetis}, because number of variables (which is $21$) is less than number of equations (which is $24$). Razaviyayn et al. \cite{Razav} proves that improper system of equations are infeasible when each transmitter uses only one beamforming vector. Therefore, using $3$ extension of the channel, we can not achieve $6$ degrees of freedom for this network. In the following, it is proved that this system of IA requirements,  (\ref{SE1}),  is infeasible using every finite extension of the channel. 

Consider a $3n$ symbol extension of the channel. Over this extended channel, the only  achievable scheme is the case  where each of the $6$ messages achieves $n$ DoF if possible,  using beamforming at every transmitter and zero-forcing at every receiver. The $n \time 1$ vectors ${\bf V}^{[j]}, j=1, \ldots, 6$ should satisfy IA conditions along with the linear independence condition.  IA requirements at $3$ receivers can be summerized as follows,  
\begin{eqnarray}
\textrm{span} \left ( {\bf H}^{[ij]} {\bf V}^{[j]}  \right )=\textrm{span} \left ( {\bf H}^{[ik]} {\bf V}^{[k]} \right ),  \quad \forall j, k \in \bar{S}_i.
\label{G6b3ia}
\end{eqnarray}

 Since diagonal channel matrices $ {\bf H}^{[ij]}$,  are full rank almost surely,   After some algebric manipulations on (\ref{SE1}), (\ref{G6b3ia}) implies that, 
 \begin{eqnarray}
\textrm{span} \left ( {\bf T}_{j, u}^{[i]}{\bf V}^{[u]}  \right )=\textrm{span} \left({\bf V}^{[u]}\right ),  \quad i=2, 3 \quad \forall u \in S_1  \cap \bar{S}_i,    \quad \forall j \in \bar{S}_1 \cap  \bar{S}_i 
\label{G6by3e2}
\end{eqnarray} 
Where ${\bf T}_{j, u}^{[i]}$'s are defined as follows,  
\begin{eqnarray}
\begin{array}{c}
 {\bf T}_{6, 1}^{[2]}={\bf T}_{6, 4}^{[2]}={\bf H}^{[26]}\left ( {\bf H}^{[16]} \right )^{-1} {\bf H}^{[13]} \left ( {\bf H}^{[23]} \right )^{-1} \\ {\bf T}_{2, 1}^{[3]}=\left ( {\bf H}^{[31]} \right )^{-1}{\bf H}^{[32]}\left ( {\bf H}^{[12]} \right )^{-1} {\bf H}^{[13]} \left ( {\bf H}^{[23]} \right )^{-1}{\bf H}^{[21]} \\
{\bf T}_{2, 4}^{[3]}=\left ( {\bf H}^{[34]} \right )^{-1}{\bf H}^{[32]}\left ( {\bf H}^{[12]} \right )^{-1} {\bf H}^{[13]} \left ( {\bf H}^{[23]} \right )^{-1}{\bf H}^{[24]} \\
{\bf T}_{5, 1}^{[3]}=\left ( {\bf H}^{[31]} \right )^{-1}{\bf H}^{[35]}\left ( {\bf H}^{[15]} \right )^{-1} {\bf H}^{[13]} \left ( {\bf H}^{[23]} \right )^{-1}{\bf H}^{[21]} \\
{\bf T}_{5, 4}^{[3]}=\left ( {\bf H}^{[34]} \right )^{-1}{\bf H}^{[35]}\left ( {\bf H}^{[15]} \right )^{-1} {\bf H}^{[13]} \left ( {\bf H}^{[23]} \right )^{-1}{\bf H}^{[24]} \end{array}
\label{63ex}
\end{eqnarray}

 (\ref{G6by3e2}) implies that there is at least one eigenvector of  ${\bf T}_{j, u}^{[i]}$ in  $\textrm{span} \left ({\bf V}^{[u]} \right ),  u \in S_1$. Since all channel matrices are diagonal,    the set of eigenvectors of channel matrices,    their inverse and product are column vectors of the identity matrix.  Define ${\bf e}_k=[0 \; 0 \; \cdots \; 1 \; \cdots \; 0]^T$ and note that ${\bf e}_k$ exists in $\textrm{span} \left (  {\bf V}^{[u]}  \right ),\; \forall u \in S_1$,  therefore, the set of equations in  (\ref{SE1}) implies that 
\begin{eqnarray}
&{}& {\bf e}_k \in \textrm{span} \left ( {\bf H}^{[ij]} {\bf V}^{[j]}  \right ),    \quad \forall (i,  j) \in \{1,  2,  3\} \times \{1,   \ldots,   6\}
\label{CAE4}           
\end{eqnarray}

Thus,    at receiver $1$,    the desired signal $ {[\bf H}^{[11]} {\bf V}^{[1]},  {\bf H}^{[14]} {\bf V}^{[4]}] $  is not linearly independent of the interference signal,    ${\bf H}^{[12]} {\bf V}^{[2]}$,    and hence,    receiver $1$ can not fully decode $W_1$ and  $W_4$ solely by zeroforcing the interference signal. Therefore,    if the channel coefficients are completely random and generic,    we can not obtain $6/3$ DoF for the $6 \times 3$ user single antenna interference channel through linear IA schemes.
\end{IEEEproof}
\section{Channel Aided IA For General Message Demands}

IA scheme,  used in  \cite{Ke},  achieves total number of DoF asymptotically when the duration of time expansion goes to infinity.  Our objective is to achieve the same performance using limited channel extensions.

Optimum DoF assignment in an interference channel with general mesage demands is obtained by solving the following linear programming problem;
\begin{eqnarray} \begin{split}
&{\bf d^*}=\rm{arg} \max_{\bf d} \; {\bf w}^T {\bf d}\\
&\quad  s.t \quad {\bf z} \preceq {\bf w}, \\ & \quad {\bf d} \succeq 0.
\end{split}
\label{dofmaxprob}
\end{eqnarray} 
Where  {\bf w} is an all one vector, ${\bf w}=[1,  \ldots,  1]^T$,  $ {\bf z}$ is defined as a $G \times 1$ vector consisted of elements $z_i=\sum_{i\in S_i} d_i + \max_{j\in \bar{S}_i} d_j$. 
The solution for each specific configuration can be obtained using methods like simplex algorithm. Although there is no closed form solution for general case of arbitrary requested message set structure,  however,  we can make some observations on the general solution.

We can assume that each of the sets  $S_i $ has at most $K-2$ elements. Otherwise,  if a set,  say $S_i $,  would have $K$ or $K-1$ elements,  receiver $i$ and its corresponding transmitters can be considered as a multiple access channel without losing any DoF gain,  and therefore,  total number of DoF of this network is $1$ and optimum DoF assignment is every vector ${\bf d}\succeq 0$ that satisfies ${\bf w}^T {\bf d}=1$. There is no need for IA in this case and simple methods like time division based multiple access techniques can achieve total number of Dof of this structure.

Assuming,  without loss of generality,  that $d_1^* \geq d_2^* \geq \cdots \geq d_K^*$,  where $d_i^*$ is the optimal value of $d_i$ obtained by solving (\ref{dofmaxprob}), we have the following theorem,  
\begin{theorem}
In an interference network with $|S_i|<K-1$,   we should have $d_1^*=d_2^*\leq \frac{1}{2}$.
\label{theo2}
\end{theorem}
\begin{IEEEproof}
 The proof can be found in appendix.

\end{IEEEproof}

\begin{corollary}
If $d_1>d_2\geq d_3 \geq \ldots \geq d_K$,  total number of DoF is obtained as $\sum_{i=1}^K d_i=1$.
\end{corollary} 
\begin{IEEEproof}
The Lagrange dual problem for (\ref{dofmaxprob}) is obtained as 
\begin{eqnarray} \begin{split}
&{\bf d}=\rm{arg} \min_{\boldsymbol{\lambda}}{\bf w}^T \boldsymbol{\lambda}\\
&\quad  s.t \quad {\bf A}^T \boldsymbol{\lambda} \preceq 1, \\ &\quad \boldsymbol{\lambda} \succeq 0.
\end{split}
\label{dofdualprob}
\end{eqnarray} 
Since strong duality holds for the optimiztion problem (\ref{dofmaxprob}),  thus,  ${\bf w}^T \boldsymbol{\lambda}^*={\bf w}^T {\bf d}^*$. On the other hand,  it is obtained in (\ref{lamb1}) that ${\bf w}^T \boldsymbol{\lambda}^*=1$,  therefore,  the proof is complete. 
\end{IEEEproof}
 
\subsection{The case of regular Interference Networks}


\begin{theorem}
\label{theox}
In a $K \times N$ user regular IC,  assuming the channel model described in section \ref{secsysmod},    the necessary and sufficient condition for perfect interference alignment to be feasible in finite channel extension is  the following structure of the channel matrices:
\begin{eqnarray}
 {\bf T}_{j, u}^{[i]}={\bf P}_{n(\beta+1)} \left [ \begin{array}{c c c} \tilde{{\bf T}}_{j, u}^{[i]} & 0 & 0 \\ 0 & \tilde{{\bf T}}_{j, u}^{[i]}& 0 \\ 0 & 0 & f(\tilde{{\bf T}}_{j, u}^{[i]}) \end{array} \right ] {\bf P}_{n(\beta+1)}^T, 
 \label{gencac}
 \end{eqnarray}

 Where  ${\bf T}_{j, u}^{[i]}$ matrices are diagoanl matrices depending on channel matrices and message demand sets structure and would be derived for each specific interference network in the following,  ${\bf P}_{n(\beta+1)}$ is an arbitrary $n(\beta+1) \times n(\beta+1)$ permutation matrix,  $\tilde{ {\bf T}}_{j, u}^{[i]} $ is an  arbitrary  $n_1 \times n_1$  diagonal matrix, $n_1$ is an arbitrary non zero integer number not greater than $n$,  and  $f({\bf X})$ is a  mapping whose domain is an arbirary $n_1 \times n_1$ diagonal  matrix and range is an $(n(\beta+1)-2n_1) \times (n(\beta+1)-2n_1) $  diagonal matrix ${\bf Y}=f({\bf X})$ whose set of diagonal elements  is a subset of diagonal elements of  ${\bf X}$.
\end{theorem}

\begin{IEEEproof}
Consider again the $6 \times 3$ interference network described in Fig. \ref{figure:6by3user},  the proof for general case is similar and is ommited here for conciseness. Considering $3n$ extension of the channel,  each receiver should achieve $2n$ DoF out of $3n$ available dimensions,  $n$ of available dimensions is assigned to interference. Theorem \ref{theox} for this special case can be written as follows:
\begin{eqnarray}
 {\bf T}_{j, u}^{[i]}={\bf P}_{3n} \left [ \begin{array}{c c c} \tilde{{\bf T}}_{j, u}^{[i]} & 0 & 0 \\ 0 & \tilde{{\bf T}}_{j, u}^{[i]}& 0 \\ 0 & 0 & f(\tilde{{\bf T}}_{j, u}^{[i]}) \end{array} \right ] {\bf P}_{3n}^T,  \quad i=2, 3 \quad \forall u \in S_1  \cap \bar{S}_i,    \quad \forall j \in \bar{S}_1 \cap  \bar{S}_i.
 \label{6by3cac}
 \end{eqnarray}
 
where ${\bf T}_{j, u}^{[i]}$ matrices are defined in (\ref{63ex}). 

\begin{lemma}
\label{lemma1}
Assuming that $ {\bf V}^{[1]}$  is of rank $n$,   (\ref{G6by3e2}) implies that   $n$  eigenvectors of ${\bf T}_{j, u}^{[i]}$ lie in $\textrm{span} \left (  {\bf V}^{[1]}  \right )$.
\end{lemma}
\begin{IEEEproof}
The proof is similar to the one present in \cite{Samadi}
\end{IEEEproof}
Based on the discussion we had on (\ref{CAE4})$, \textrm{span} \left ( {\bf V}^{[1]}  \right )$ should not contain any vector of the form ${\bf e}_i$,   and since $\textrm{span} \left ( {\bf V}^{[1]}  \right )$ has dimension $n$,   it should have $n$ basis vectors of the form $ {\bf v}\tilde{{\bf T}}=\sum_{i=1}^{2n}\alpha_i {\bf e}_i,   \quad j=1, \ldots,  n$,   where at least $2$ of $\alpha_i$'s are nonzero. Let's call vectors with this form as non ${\bf e}_i$ vectors. Since $n$ of ${\bf T}_{j, u}^{[i]}$'s eigenvectors lie in $\textrm{span} \left ( {\bf V}^{[1]}  \right )$,   the matrix ${\bf T}_{j, u}^{[i]}$ should have at least $n$ non ${\bf e}_i $ eigenvectors. Note that this requirement is necessary not sufficient. Assuming that ${\bf S}=[{\bf s}]$ is a matrix consisted of non ${\bf e}_i $ eigenvectors of ${\bf T}_{j, u}^{[i]}$ as its columns, it is concluded that $\textrm{span} \left ( {\bf V}^{[1]}  \right ) \in \textrm{span} \left ( {\bf S}  \right )$.

\begin{lemma}
\label{lemmker}
${\bf T}_{j, u}^{[i]}$ has no unique diagonal element.
\end{lemma}
\begin{IEEEproof}
It is easy to see that if ${\bf s}_1= {\bf e}_p + {\bf e}_q,   \quad p,q=1,    \ldots,   n, p \neq q$  is an eigenvector of ${\bf T}_{j, u}^{[i]}$,   then ${\bf T}_{j, u}^{[i]}(p) ={\bf T}_{j, u}^{[i]}(q)$.   If ${\bf T}_{j, u}^{[i]}(p) $ is unique,  this implies that non ${\bf e}_p $ eigenvectors of ${\bf T}_{j, u}^{[i]}$ do not contain ${\bf e}_p$, and hence,  ${\bf e}_p \in \textrm{kernel} \left ( {\bf S}  \right )$, where $ \textrm{kernel} \left ( {\bf S}  \right )$  denotes  the null space of columns of matrix ${\bf S}$. Thus, $ {\bf e}_p \in \textrm{kernel} \left ({\bf V}^{[1]} \right )$ because  $\textrm{span} \left ( {\bf V}^{[1]}  \right ) \in \textrm{span} \left ( {\bf S}  \right )$. Since all channel matrices are diagonal,  using (\ref{SE1}),  ${\bf e}_p \in \textrm{kernel}({\bf V}^{[1]})$ implies that 
 \begin{eqnarray}
&{}& {\bf e}_p \in \textrm{kernel} \left ( {\bf H}^{[ij]} {\bf V}^{[j]}  \right ),    \quad \forall (i, j) \in \{1,   2,   3\} \times \{1,   \ldots,   6\}.
\end{eqnarray}

Thus,    at receiver $1$,    the total dimension of the desired signals $ [{\bf H}^{[11]} {\bf V}^{[1]},  {\bf H}^{[14]} {\bf V}^{[4]}] $ plus interference from undesired transmitters is less than $3n$,  and desired signals are not linearly independent of the interference signals,   ${\bf H}^{[1j]} {\bf V}^{[j]},  j\in \bar{S}_1$,    and hence,    receiver $1$ can not fully decode $W_1$ solely by zeroforcing the interference signal. 

Note that all $6$  channel aiding conditions  in (G6by3e2) share the same permutation matrix ${\bf P}$ and mapping function $f({\bf X})$. This is because the diagonal matrices  ${\bf T}_{j, u}^{[i]}$ should have the same set of non ${\bf e}_i$ eigenvectors which are supposed to be columns of user $1$ beamforming matrix, ${\bf V}^{[1]}$
\end{IEEEproof}
 
 Lemma \ref{lemmker} concludes the proof of the necessary part of Theorem \ref{6by3cac}. The sufficient part is easily proved by noting the fact that the matrices ${\bf T}_{j, u}^{[i]} $ with the form given in (\ref{6by3cac}) have  $L \geq n$ non ${\bf e}_i $  common eigenvectors ${\bf r}_i, i=1, \ldots, L$ with the property that 
 \begin{eqnarray}
{\bf e}_k \not \in \textrm{span}({\bf R}), \quad k=1, \ldots, 2n,
\label{spnprp}
\end{eqnarray}
 and
 \begin{eqnarray}
  {\bf e}_k \not \in \textrm{kernell}({\bf R}), \quad k=1, \ldots, 2n,
  \label{krnlprp}
  \end{eqnarray}
where ${\bf R} $  is defined as a $2n \times L$ matrix consisted of ${\bf r}_i$'s as its columns. Every $n$ subset of these eigenvectors can be considered as the columns of user $1$ transmit beamforming matrix ${\bf V}^{[1]}$. ${\bf V}^{[2]}$--${\bf V}^{[6]}$ can be designed using (\ref{SE1}).
\end{IEEEproof}

\begin{example}
Considering $3$ extension of the channel. Since $\beta=2$ for this example structure, hence, $n=2$ and $n_2$, and the following simplified channel aiding condition is derived, 
 \begin{eqnarray}
 {\bf T}_{j, u}^{[i]}=\kappa_{j, u}^{[i]}  {\bf I}_{3},  \quad i=2, 3 \quad \forall u \in S_1  \cap \bar{S}_i,    \quad \forall j \in \bar{S}_1 \cap  \bar{S}_i 
 \label{excac1}
 \end{eqnarray}
 where $\kappa_{j, u}^{[i]}$ is a nonzero arbitrary number. In fact, (\ref{G6by3e2}) implies that ${\bf V}^{[1]}$ should be designed to be an eigenvector of  ${\bf T}_{j, 1}^{[i]}$. At the same time,   ${\bf V}^{[1]}$,  based on the discussion on (\ref{CAE4}),  should satisfy the following condition
 
 \begin{eqnarray}
{\bf e}_i \not \in \textrm{span}({\bf V}^{[1]}), i=1, 2, 3,  
\label{spancond}
\end{eqnarray}
   which, in this case, simply means that ${\bf V}^{[1]}$ should not be a multiple of ${\bf e}_k$.  Therefore ${\bf V}^{[1]}$ can be written in the following form 
\begin{eqnarray}
{\bf V}^{[1]}=\sum_{i=1}^3 \alpha_i {\bf e}_i,  
\label{vvec}
\end{eqnarray}
 where at least two of $\alpha_i$'s are nonzero.On the other hand, Lemma \ref{lemmker} implies that  all $\alpha_i'$s  in (\ref{vvec}) should be nonzero.  Considering the fact that  ${\bf T}_{j, u}^{[i]}$ is a diagonal matrice,  (\ref{G6by3e2}) and (\ref{vvec}) imply that ${\bf T}_{j, u}^{[i]}$ should have an eigenvector of the form  (\ref{vvec}),  which means that ${\bf T}_{j, u}^{[i]}$ has three equal nonzero diagonal elements,  i.e., 
 \begin{eqnarray}
 {\bf T}_{j, u}^{[i]}=\kappa_{j, u}^{[i]}  {\bf I}_{3},  \quad i=2, 3 \quad \forall u \in S_1  \cap \bar{S}_i,    \quad \forall j \in \bar{S}_1 \cap  \bar{S}_i 
 \label{excac}
 \end{eqnarray}
 
 Therefore, ${\bf V}^{[u]},  u \in S_1$ can be designed arbitrarily as long as it contains no zero element. 
\end{example}
\begin{remark}
It can be noticed in (\ref{63ex}) that 
 \begin{eqnarray}
 {\bf T}_{5, 4}^{[3]}={\bf T}_{2, 4}^{[3]}\left ({\bf T}_{2, 1}^{[3]} \right )^{-1}{\bf T}_{5, 1}^{[3]}, 
\end{eqnarray}
thus,  the channel aiding condition related to this matrix,  $ {\bf T}_{5, 4}^{[3]}=\kappa_{5, 4}^{[3]}  {\bf I}_{3}$,   is already satisfied and this condition does not contribute to a new channel aiding condition. Overall,  in this example interference network,  we have $4$ independent channel aiding conditions for interference signals to be aligned perfectly at all receivers. 
\end{remark}
 
 \begin{remark}
 Note that the channel can be extended in frequency (e.g., the OFDM channel) or time domain. If the channel is extended in time domain,  $n(\beta+1)$ used time slots,  which satisfy channel aiding conditions,  are not necessarily succesive.
 \end{remark}
 
\begin{remark}
Consider the special case of $K$ user interference channel,  which is analysed in detail in \cite{Samadi}. In this case $S_i={i}$, and we can drop index $u$ in (\ref{gencac}),  because $S_1={1}$ consists of a single element. Channel aiding condition in this case can be obtained from  (\ref{gencac}) as follows, 
\begin{eqnarray}
  {\bf T}_{j}^{[i]}= {\bf P}_{2n}  \left [ \begin{array}{c c c} \tilde{{\bf T}}_{j}^{[i]} & 0 & 0 \\ 0 & \tilde{{\bf T}}_{j}^{[i]}& 0 \\ 0 & 0 & f(\tilde{{\bf T}}_{j}^{[i]}) \end{array} \right ] {\bf P}_{2n}^T, \quad  i,  j\in \{2, 3, \cdots, K\},  \quad i\neq j,  
\label{CAKm}
\end{eqnarray}
where ${T}_{j}^{[i]}$ matrices are defined as 
\begin{eqnarray}
{\bf T}_{j}^{[i]}=\left ( {\bf H}^{[i1]} \right )^{-1} {\bf H}^{[ij]} \left ( {\bf H}^{[1j]} \right )^{-1} {\bf H}^{[13]} \left ( {\bf H}^{[23]} \right )^{-1}{\bf H}^{[21]}, \quad  i,  j\in \{2, 3, \cdots, K\},  \quad i\neq j.
\end{eqnarray}
\end{remark}
\begin{remark}
The assumption that all of the receivers requests the same number of transmitted symbols,  and each of the messages is requested by the same number of prime receivers is presumed because achievable scheme for this case is equal DoF assignment for all transmitters. Every other network structure that achieves its total number of DoF by assigning zero or equal number of DoF to each transmitter can utilize the scheme proposed in this section. As an example,   consider an interference channel with four transmitters and three prime receivers,  the message request sets are $\{1,  2\}$, $\{1, 3\}$,  $\{1,  4\}$. Optimal DoF assignment can be obtained by solving the following linear programming problem
\begin{eqnarray} \begin{split}
&(d_1,  d_2,  d_3,  d_4)=\rm{arg} \max_{d_i} \sum_{i=1}^4 d_i \\
&\quad \quad s.t \quad  \sum_{i\in S_i} d_i + \max_{j\in \bar{S}_i} d_j \leq 1,    \\
& \quad \quad d_i \geq 0,  \quad \forall i=\{1,  \ldots,  4\}. 
\end{split}
\end{eqnarray}
Which is obtained to be $(0,  \frac{1}{2},  \frac{1}{2},  \frac{1}{2})$. Channel aided IA can be applied to this network by excluding transmitter $1$ and using $2$ extension of the channel to deliver remaining transmitted message.
\end{remark}

\subsection{The case of irregular interference networks}

In this section, we will develop an algorithm to obtain suffiecient channel aiding conditions to achieve perfect IA for irregular interference networks. Assume that the optimum DoF assignment is in the form of $(d_1,  d_2, \ldots,  d_J)$,  which are not equal in general. $J \leq K$ is the number of active users,  i.e., $d_i>0,  \forall i=1,  \ldots,  J$. These DoF assignments achieve total number of DoF of the interference channel with general message demands. Since all coefficients and right hand side bounds of the maximization problem (\ref{dofmaxprob}) are integers,  optimal DoF assignments are rational. Consider $N_e$ extension of the channel where $N_e$ is an integer number such that $N_e d_j \in \mathrm{Z}^{+},  \forall j=1,  \ldots,  J$. Define $d^{0}_j=N_ed_j$ which is an integer number. Using $N_e$ extension of the channel,  it is evident that ${\bf d^{0}} =\{d^{0}_1,  \ldots,  d^{0}_J\}$ achieves total number of DoF of the channel. We investigate channel aiding conditions in this case. 

Without loss of generality,  we can assume that $d^{0}_1= d^{0}_2 \geq \cdots \geq d^{0}_J$,  
\begin{enumerate}
\item
Consider an interference network with $J_0=J$  transmitters and $N$ receivers along with the sets derived from active users $S_{i}^0=S_{i},  \bar{S}_{i}^0= \bar{S}_{i}$. Consider the first $d^s_i=d^0_{J}$ columns of all transmitter precoding matrices. Constitute the new sets of requestet meassages $S_{i}^s$ and interfering meassages  $\bar{S}_{i}^s$,  consisting of transmitters with the number of transmitted messages being greater than zero. Channel aiding conditions required to  perfectly align receivied interference from these set of  transmitted messages  at each receiver within $d^s_i$ dimensions of $N_e$ available dimensions are derived as follows,  
 \begin{eqnarray} \begin{split}
  {\bf T}_{j, u}^{[i]}= {\bf P}_{N_e}  \left [ \begin{array}{c c c} \tilde{{\bf T}}_{j, u}^{[i]} & 0 & 0 \\ 0 & \tilde{{\bf T}}_{j, u}^{[i]}& 0 \\ 0 & 0 & f(\tilde{{\bf T}}_{j, u}^{[i]}) \end{array} \right ]{\bf P}_{N_e}^T,   \quad \\   i\in \{2, 3, \cdots, N\},  \quad \forall u \in S_1^s,  \quad  \forall j \in \bar{S}_1^s \cap  \bar{S}_i^s \end{split} 
 \label{gencac2}
 \end{eqnarray}
where $\tilde{{\bf T}}_{j, u}^{[i]}$ is an arbitrary diagonal matrix, and $f(\tilde{{\bf T}}_{j, u}^{[i]})$ is the same mapping  defined for (\ref{gencac}). We should remind that the mapping $f(\tilde{{\bf T}}_{j, u}^{[i]})$ is used to make sure that ${\bf e}_k \not \in \rm{span}(V^{[l]}),  \forall k \in [N_e],  l \in [J]$. The property $d^0_{min} \leq \frac{N_e}{2}$ makes sure that this set of conditions are feasible. 
\item 
Consider a new interference network with $J_1$ transmitters,  and $N$ receivers,  where $J_1$ is defined as the number of transmitters with remaining number of DoF $d_i^1=d_i^0-d_i^s$ being greater than zero. Denote respective remaining transmitted message sets  $S_{i}^1 \; \textrm{and} \; \bar{S}_{i}^1$ Repeat step $1$ for this new network structure. Note that $d^1_{J_1} \leq \frac{N_e}{2}$ is still valid. 

This procedure is repeated until all transmitters send their respective messages.  Derived channel aiding conditions are sufficient feasibility  conditions for perfect IA. Precoding matrices at each step can be designed as described before. 
\end{enumerate}
\begin{example}
Consider an example case of a $5 \times 3$ interference channel with requested message sets defined as $S_1=\{1,  5\},  S_2=\{1, 2\} \; \textrm{and} \; S_3=\{3,  4,  5\}$. Solving the linear programming (\ref{dofmaxprob}),  optimum DoF assignments are obtained as $d_1=d_2=0.4,  d_3=d_4=d_5=0.2$,  and total number of DoF is obtained as $\sum_{i=1}^5 d_i=1.4$. Consider $5$ extension of the channel,  each of  transmitters $1$ and $2$ should send $2$ independent messages and tranmitters $3,  4, \; \textrm{and} \; 5$ each sends an inpendent message. 
\begin{figure}
\centering \includegraphics[scale=.6]{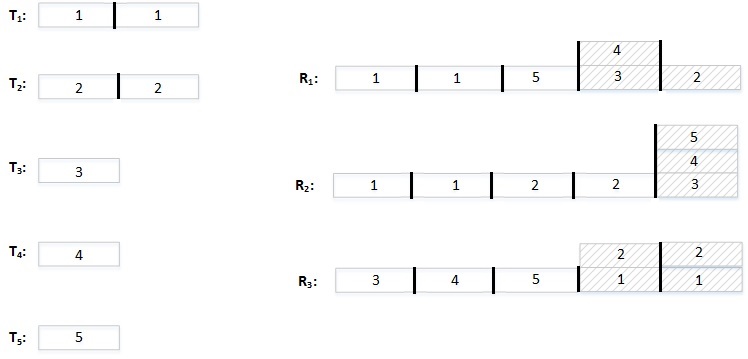}
\caption{IA strategy for a $5 \times 3$ interference channel,  dashed rectangles represent interfering messages.}
\label{fig:GIC}
\end{figure}
Alignment strategy for this structure is shown in Fig. \ref{fig:GIC}. Consider the first column of each transmitter precoding matrix,  i.e., each transmitter sends a single message, $d^s_i=1, i\in \{1, \ldots, 5\}$.  The transmitting set of requested message sets are $S^s_1=\{1,  5\}, S^s_2=\{1,  2\} \; \textrm{and}\; S^s_3=\{3,  4,  5\}$ and transmitted interfering message sets are $\bar{S}^s_1=\{2,  3,  4\},  \bar{S}^s_2=\{3,  4,  5\}  \; \textrm{and}\;  \bar{S}^s_3=\{1,  2\}$. The only Channel aiding condition to  perfectly align receivied interference from these transmitted messages  at all receivers within $1$ dimension out of $5$ available dimensions, considering described network structure,  is derived as follows,  
 \begin{eqnarray}
  {\bf T}^s=  \eta {\bf I}_{5},  
 \end{eqnarray}
where $ \eta $ is an arbitrary nonzero constant number and  ${\bf T}^s$ is defined as
 \begin{eqnarray}
  {\bf T}^s=\left ( {\bf H}^{[23]} \right )^{-1} {\bf H}^{[24]} \left ( {\bf H}^{[14]} \right )^{-1}{\bf H}^{[13]}.
 \label{exuneqcac}
 \end{eqnarray}
Transmitters $1$ and $2$ are the only transmitters with remaining DoFs,  $d^1_1=d^1_2=1$,  greater than zero. Setting  ${\bf d}^s=\{1,  1\}$,  the new set of requested message sets are $S^1_1=\{1\}, S^2_2=\{1,  2\} \; \textrm{and}\; S^1_3=\emptyset$ interfering message sets are $\bar{S}^1_1=\{2\},  \bar{S}^2_2=\emptyset  \; \textrm{and}\;  \bar{S}^1_3=\{1,  2\}$ which do not contribute to any new channel aiding condition. Therefore,  (\ref{exuneqcac}) is the only channel aiding condition for this structure to achieve perfect IA over limited number of channel extension. 
\end{example}

\begin{remark}
Our scheme relies on matching up certain channel matrices so that the interference terms cancel out when received signal vectors are combined linearly. Clearly,   given any matrix $ {\bf T}_{j, u}^{[i]}$,    the probability that channel aiding condition will occur exactly  is zero,  assuming continuous-valued fading. Thus,    we can only look for channel aiding conditions to be satisfied approximately. By taking finer approximations,    we can achieve the target rate in the limit. 
\end{remark}

\section {Conclusion}
The concept of regular interference network is introduced as an interference network  where all active  transmitters have equal optimal number of DoF. Interference networks where all transmitters emit messages to an equal number of receivers and all receivers request an equal number of messages are one of special cases of regular interference networks. It is proved that perfect IA can not be achieved for regular interfererence channels with generic channel coefficients. Perfect IA feasibility conditions on channel structure was addressed and an alignment scheme was introduced  to achieve total number of DoF of the interference network with general message demands,  using  limited number of channel extension. 

  Derived channel aiding conditions are necessary and sufficient for the case of regular interference networks,  and are sufficient conditions for irregular interference networks.  Overall,    the proposed method aims at reducing the required dimensionality and signal to noise ratio for exploiting DoF benefits of IA schemes.

\bibliographystyle{IEEEtran}

\section{Appendices}{Proof of the Theorem \ref{theo2}} \label{app1}

Adding the new constraint $d_1 \geq d_2 \geq \cdots \geq d_K$ to  (\ref{dofmaxprob}),  and introducing Lagrange multipliers $\boldsymbol{\lambda} \in \mathrm{R}^G$ for the inequality constraints ${\bf z} \leq 1$,  multipliers  $\boldsymbol{ \gamma} \in \mathrm{R}^K$  for the inequality constraints ${\bf d} \succeq 0$,  and multipliers  $\boldsymbol{\eta} \in \mathrm{R}^{K}$  for the inequality constraints $d_j \leq d_1,  j=1,  \ldots,  K$,  we obtain the following KKT conditions,  
\begin{enumerate}
\item ${\bf d}^* \succeq 0,\;  {\bf z}^* \preceq {\bf w},\;  d_1^* \geq d_2^* \geq \cdots \geq d_K^*$
\item $\boldsymbol{\gamma}^* \succeq 0, \quad  \boldsymbol{\lambda}^* \succeq 0,  \quad\boldsymbol{ \eta}^* \succeq 0 $
\item $d_i^* \gamma_i^*=0,\; (d_i-d_1)^* \eta_i^*=0,\; \forall i=\{1, \ldots, K\}$,\;  $z_j^* \lambda_j^*=0,\; \forall j=\{1, \ldots, G\}$, 
\item $\bigtriangleup (L({\bf d},  \boldsymbol{\lambda,  \gamma,  \eta} ))|_{{\bf d}^*, \boldsymbol{\lambda}^*, \boldsymbol{ \gamma}^*,  \boldsymbol{\eta}^*}=0$.
\end{enumerate}
Where $(\cdots)^*$ superscript indicates the respective optimal values of primal and dual optimization problems, and $L({\bf d},  \boldsymbol{\lambda,  \gamma,  \eta} )$ is the lagrangian function. 
\begin{eqnarray}
L({\bf d},  \boldsymbol{\lambda,  \gamma,  \eta} )= -{\bf w}^T {\bf d}+  \boldsymbol{\lambda}^T({\bf z}-{\bf w})-\boldsymbol{ \gamma}^T{\bf d}+ \boldsymbol{\eta}^T({\bf d}-d_1{\bf w})
\end{eqnarray}

Assume that $d_1^*$ is the unique maximum DoF number,  i.e $d_1^*>d_j^*,  j=2,  \ldots,  K$,  evaluating derivation  with respect to $d_1$ in  $4^{th}$ KKT condition,  we obtain,  
\begin{eqnarray}
-1+\sum_{i=1}^G \lambda^*_i -\gamma^*_1-\sum_{j=1}^{K-1} \eta^*_j=0.
\label{derto1}
\end{eqnarray}
Since $d_1^*>0$ and $d_1^*>d_j^*$,  complementary slackness conditions imply that $\gamma_1^*=0$ and $\eta_j^*=0,  j=1,  \ldots,  K-1$. Substituting these values into  (\ref{derto1}),  it is obtained that 
\begin{eqnarray}
\sum_{i=1}^G \lambda_i^* =1.
\label{lamb1}
\end{eqnarray}
In general,  evaluating $4'$th KKT condition with respect to $d_i,  i=1,  \ldots,  K$ and noting the fact that $\eta_j^*=0$,  we get
\begin{eqnarray}
{\bf a}_i^T \boldsymbol{\lambda}^* =1+\gamma_i.
\label{lambi}
\end{eqnarray}
Where ${\bf a}_i$ is difined as a $G \times 1$ vector consisting of $0$ or $1$ elements obtained by taking derivaties ${\bf a}_i(j)=\frac{\partial z_j}{\partial d_i},  (i, j) \in \{1,  \ldots,  K\} \times \{1,  \ldots,  G\}$. Since $\boldsymbol{\lambda}^* \succeq 0$,  hence,   ${\bf a}_i^T \boldsymbol{\lambda}^* \leq {\bf w}^T \boldsymbol{\lambda}^*=1$,  therefore $\gamma_i=0$. Suming (\ref{lambi}) over $i$,  we obtain
\begin{eqnarray}
&\sum_{i=1}^K {\bf a}_i^T \boldsymbol{\lambda}^* =K \nonumber \\
&\quad \Rightarrow \sum_{j=1}^G (\beta_j+1) \lambda_j=K
\label{lambsum}
\end{eqnarray}
(\ref{lambsum}) is not feasible unless $\beta_j \geq K-1,  j=1,  \ldots,  G$,  this contradicts the assumption that each receiver requests less than $K-1$ transmitted messages. Therefore,  if we have more than two prime receivers,  the set $d_1^*\geq d_2^* \geq  \ldots \geq d_K$ can not have a unique maximum and $d_1^*=d_2^*$. 

Considering a set $S_{J0}$ for which $d_1^*\in S_{J0}$,   DoF region inequality for this set  $z_{j0}\leq 1$ implies that 
\begin{eqnarray}
z_{j0}\leq 1 \Rightarrow d_1^* +d_2^* \leq 1 \Rightarrow  d_i^* \leq \frac{1}{2},  \forall i=1,  \ldots,  K. 
\end{eqnarray}

\end{document}